\documentclass[a4paper,fleqn,12pt]{cas-sc}

\usepackage[numbers]{natbib}
\usepackage{color,soul}
\usepackage[normalem]{ulem}
\usepackage{float} 
\usepackage{bm}
\restylefloat{table}
\restylefloat*{figure}
\usepackage{graphics}
\usepackage{caption}
\usepackage{amsmath}
\usepackage{amssymb}
\usepackage{amsthm}
\usepackage{amsfonts}
\usepackage{epsfig} % Include figure files
\usepackage{graphicx}% Include figure files
\usepackage{dcolumn}% Align table columns on decimal point
\usepackage{setspace}
\usepackage{comment}
\usepackage{lineno}
\usepackage{adjustbox}
\usepackage{appendix}
\usepackage{subcaption}
\usepackage[dvipsnames]{xcolor}

\usepackage{booktabs}
\usepackage{multirow,ulem}
\usepackage{hhline,colortbl,makecell}

\definecolor{drkY}{rgb}{1.0, 0.83, 0.0}
\definecolor{ltY}{rgb}{0.99, 0.99, 0.59}
\definecolor{drkT}{rgb}{0.0, 0.81, 0.82}
\definecolor{ltT}{rgb}{0.69, 0.93, 0.93}
\definecolor{drkR}{rgb}{1.0, 0.25, 0.25}
\definecolor{ltR}{rgb}{1.0, 0.5, 0.47}
\definecolor{ltG}{rgb}{0.7, 0.93, 0.36}
\definecolor{lgray}{rgb}{0.83, 0.83, 0.83}
\definecolor{blk}{rgb}{0.0, 0.0, 0.0}
\definecolor{txtR}{rgb}{0.89, 0.0, 0.13}
\definecolor{txtG}{rgb}{0.0, 0.42, 0.24}

\let\printorcid\relax

\usepackage{geometry}
\begin{document}

%%%%%%%%%%%%%%%%%%%%%
% Line numbering
%%%%%%%%%%%%%%%%%%%%%
%\usepackage{lineno}
% Please use line numbering with your initial submission and
% subsequent revisions. After acceptance, please turn line numbering
% off by adding percent signs to the lines %\usepackage{lineno} and
% to %\linenumbers{} and %\modulolinenumbers[3] below.

\let\WriteBookmarks\relax
\def\floatpagepagefraction{1}
\def\textpagefraction{.001}
\shorttitle{Stability and Fluctuations in Complex Ecological Systems}
%\shortauthors{Forgoston, Day, Doelman, Hartemink, Hastings, Hemerik, Hening, Hofbauer, K{\'e}fi, Kessler, Klauschies, Kuehn, Li, Moore, Morri{\"e}n, Neutel, Pantel, de Ruiter, Schreiber, Shaw, Shnerb, Siero, Storch, Strickler, Thorne, van de Leemput, van Velzen and Weinans }

\title[mode=alt]{\LARGE Stability and Fluctuations in Complex Ecological Systems}    
%\tnotemark[1,2]

%\tnotetext[1]{We gratefully acknowledge the support of ONR Award No. 14-19-1-2253 and NSF DUE Award 1839686.}

\author[1]{\color{black}Eric Forgoston}
%\cormark[1]
%\ead{eric.forgoston@montclair.edu }
\address[1]{Department of Applied Mathematics and Statistics, Montclair State University, Montclair, NJ 07043, USA}

\author[2]{\color{black}Sarah Day}
\address[2]{Department of Mathematics, William \& Mary, Williamsburg, VA 23187, USA}

\author[3]{\color{black}Peter C. de~Ruiter}
\address[3]{Biometris, Wageningen University and Research Centre, 6708 PB Wageningen, The Netherlands and Institute of Biodiversity and Ecosystem Dynamics, University of Amsterdam, 1098 XH Amsterdam, The Netherlands
}

\author[4]{\color{black}Arjen Doelman}
\address[4]{Mathematical Institute, Leiden University, 2300 RA, Leiden, The Netherlands}

\author[5]{\color{black}Nienke Hartemink}
\address[5]{Biometris, Wageningen University and Research Centre, 6708 PB Wageningen, The Netherlands
}

\author[6]{\color{black}Alan Hastings}
\address[6]{Department of Environmental Science and Policy, University of California, Davis, CA 95616, USA and Santa Fe Institute, 1399 Hyde Park Road, Santa Fe, NM 87501, USA
}

\author[5]{\color{black}Lia Hemerik}
%\address[5]{Biometris, Wageningen University and Research Centre, 6708 PB Wageningen, The Netherlands
%}

\author[7]{\color{black}Alexandru Hening}
\address[7]{Department of Mathematics, Texas A \& M University, College Station, TX 77843, USA
}

\author[8]{\color{black}Josef Hofbauer}
\address[8]{Institute for Mathematics, University of Vienna, Oskar-Morgenstern-Platz 1, 1090 Vienna, Austria
}

\author[9]{\color{black}Sonia K{\'e}fi}
\address[9]{ISEM, CNRS, Universit{\'e} Montpellier, IRD, EPHE, Montpellier, France
}

\author[10]{\color{black}David A. Kessler}
\address[10]{Department of Physics, Bar-Ilan University, Ramat-Gan 52900, Israel
}

\author[11]{\color{black}Toni Klauschies}
\address[11]{Department of Ecology and Ecosystem Modelling, Institute of Biochemistry and Biology, University of Potsdam, Potsdam, Germany
}

\author[12]{\color{black}Christian Kuehn}
\address[12]{Department of Mathematics, Technical University of Munich, Boltzmannstr. 3, 85748 Garching b. München, Germany and Complexity Science Hub Vienna, Josefstädter Str. 39, 1080 Vienna, Austria
}

\author[13]{\color{black}Xiaoxiao Li}
\address[13]{School of Ecology, Environment and Resources, Guangdong University of Technology, Guangzhou 510006, China
}

\author[14]{\color{black}John C. Moore}
\address[14]{Natural Resource Ecology Laboratory, Colorado State University, Fort Collins, CO 80524, USA and Department of Ecosystem Science and Sustainability, Colorado State University, Fort Collins, CO 80524, USA
}

\author[15]{\color{black}Elly Morri{\"e}n}
\address[15]{Institute for Biodiversity and Ecosystem Dynamics, Earth Surface Sciences Group (IBED-ESS), University of Amsterdam, PO Box 94246, Amsterdam 1090 GE, The Netherlands
}

\author[16]{\color{black}Anje-Margriet Neutel}
\address[16]{British Antarctic Survey, Cambridge, CB3 0ET, UK
}

\author[17]{\color{black}Jelena Pantel}
\address[17]{Ecological Modelling, Faculty of Biology, University of Duisburg-Essen, Essen, Germany
}

\author[18]{\color{black}Sebastian J. Schreiber}
\address[18]{Department of Evolution and Ecology, and Center for Population Biology, University of California, Davis, CA 95616, USA
}

\author[2]{\color{black}Leah B. Shaw}
%\address[2]{Department of Mathematics, William \& Mary, Williamsburg, VA 23187, USA}

\author[10]{\color{black}Nadav Shnerb}
%\address[10]{Department of Physics, Bar-Ilan University, Ramat-Gan 52900, Israel
%}

\author[5]{\color{black}Eric Siero}
%\address[5]{Biometris, Wageningen University and Research Centre, 6708 PB Wageningen, The Netherlands
%}

\author[19]{\color{black}Laura S. Storch}
\address[19]{Mathematics Department, Bates College, Lewiston, ME 04240, USA
}

%\author[20]{\color{black}Edouard Strickler}
%\address[20]{Universit{\'e} de Lorraine, CNRS, Inria, IECL, Nancy, France
%}

\author[16]{\color{black}Michael A.S. Thorne}
%\address[16]{British Antarctic Survey, Cambridge, CB3 0ET, UK
%}

\author[20]{\color{black}Ingrid van~de~Leemput}
\address[20]{Department of Environmental Sciences, Wageningen University, 6700 AA Wageningen, The Netherlands
}

\author[11]{\color{black}Ellen van~Velzen}
%\address[11]{Department of Ecology and Ecosystem Modelling, Institute of Biochemistry and Biology, University of Potsdam, Potsdam, Germany
%}

\author[21]{\color{black}Els Weinans}
\address[21]{Copernicus Institute of Sustainable Development, Utrecht University, 3508 TC Utrecht, The Netherlands
}


\begin{abstract}
From 08--12 August, 2022, 32 individuals participated in a workshop, Stability and Fluctuations in Complex Ecological Systems, at the Lorentz Center, located in Leiden, The Netherlands. An interdisciplinary dialogue between ecologists, mathematicians, and physicists provided a foundation of important problems to consider over the next 5-10 years. This paper outlines eight areas including (1) improving our understanding of the effect of scale, both temporal and spatial, for both deterministic and stochastic problems; (2) clarifying the different terminologies and definitions used in different scientific fields; (3) developing a comprehensive set of data analysis techniques arising from different fields but which can be used together to improve our understanding of existing data sets; (4) having theoreticians/computational scientists collaborate closely with empirical ecologists to determine what new data should be collected; (5) improving our knowledge of how to protect and/or restore ecosystems; (6) incorporating socio-economic effects into models of ecosystems; (7) improving our understanding of the role of deterministic and stochastic fluctuations; (8) studying the current state of biodiversity at the functional level, taxa level and genome level.
\end{abstract}

% \includegraphics{figs/grabs.pdf}

% \begin{highlights}
% \item Research highlights item 1
% \item Research highlights item 2
% \item Research highlights item 3
% \end{highlights}


\maketitle

%\linenumbers{}
%\modulolinenumbers[3]

\section{Introduction}
One of the most well-known scientific debates occurs at the interface of Biology, Mathematics, and Physics, and relates to determining the patterns and processes that govern the stability of ecological networks. This debate originates from the middle of the last century when Robert MacArthur argued that species-rich, biodiverse communities are more likely to be stable than species-poor, less diverse communities \cite{macarthur1955}. The  argument's foundation is that diverse communities are more robust to environmental disturbances because the more complex network of ecological interactions will act as a buffer against the disturbances. This notion was well-received as it provided an interesting explanation of the large biodiversity observed in most ecosystems. However, the concept was challenged by Mark Gardner and W. Ross Ashby \cite{gardner1970}, and extended formally by Robert May \cite{may1972,may2019}, with their differing approaches suggesting the opposite: the more species and interactions that were included in the modelled communities, the less likely that these modelled communities will be stable.
     
Fifty years later, we still do not fully understand the diversity-stability relationship, despite the fact that the diversity-stability relationship has become a major debate in ecological literature. Within this debate, roughly two lines of approach have been developed. The first approach involves the inclusion of more realism in community models, in terms of deriving model parameters from biological observation. These studies show that biologically realistic models can enable the removal of the stability constraints, but a clear answer regarding how biodiversity relates to stability is still lacking. The second approach involves a more detailed mathematical analysis of the models %, particularly Jacobian matrix models, 
 to investigate the model properties that are important to stability. Alas, this approach also has not provided a definite answer to the diversity-stability issue. 

At present, the state-of-the-art in this field of research can be summarized as follows: (1) when the network structure and parameters in the mathematical models depicting the communities (or even the community matrices) are based on biological observation, then the system becomes remarkably stable \cite{deRuiter1995,jacquet2016}; and (2) analysis of a series of such empirically-based community matrices which capture a variety of levels of species richness shows that species richness does not necessarily affect the system's stability \cite{dougoud2018,jacquet2016,neutel2014}. These notions allow one to focus the diversity-stability issue towards understanding the mathematical and biological mechanisms underlying the empirically-based community matrices. 

Networks can serve as useful tools in analyzing changes in community structure over space and time. Until now, most community models of ecological networks have been based on the assumption that all species are together, always and everywhere, and hence all interactions are always realized. In real ecosystems, this is obviously not the case as groups can be temporally and spatially separated and links will then not be realized. In fact, external disturbances can lead to reduced connectivity, while restoration efforts can increase connectivity. Including such features will likely fundamentally change our understanding of flux rates and ecological functioning of food webs \cite{wang2019}, and will likely shed new light on the diversity-stability debate. 

Another relatively new research approach involves fluctuations, both deterministic and stochastic, as they likely play a key role in ecosystems.  Sources of stochasticity (random noise) can be internal, such as organisms producing a variable number of offspring, or external, such as environmental fluctuations.   Deterministic fluctuations, such as seasonality, also affect extinction and coexistence, as well as invasion and colonization (or recolonization). Although it is known that environmental noise can induce rare switching/extinction events for a variety of systems in population biology, relatively few theoretical ecological studies have considered the aspect of noise or stochasticity. Even fewer studies have considered the effects of deterministic seasonality combined with stochasticity \cite{billings2018,picoche2019} although these two aspects can effectively be studied using a very similar mathematical framework.

A major advantage in considering stochastic models is due to the fact that classical, deterministic models are often unable to capture specific ecological dynamics observed in nature.  Deterministic models are based on the mean behavior of individuals and do not account for demographic stochasticity, or the changes in population growth rates or interaction rates related to random events.  As one example, the inclusion of stochasticity through consideration of random encounters between individuals may lead to a species extinction, even when the species is fluctuating around a carrying capacity above an Allee threshold.  The inclusion of environmental stochasticity can similarly lead to observed dynamics that are not captured in the deterministic models \cite{forgoston2018}. While the  ability  to  generate stochastic  simulations  that  provide  quantitative  statistics  for  the  emergence  of  new dynamics  is  increasing  with  advances  in  computational  power,  there  remains  a  need  for  new  methods  to analyze the underlying stochastic models. Another critical issue is whether observed transitions in the frequency and
amplitude of population fluctuations (e.g., intermittent cycles) should be assigned to stochastic or environmental variability, or whether they may indicate shifts in intrinsic properties of the interacting populations \cite{vanVelzen2022}.

Ecological dynamics in (deterministically and stochastically) fluctuating environments have posed some serious challenges to analysis, but over the years, the understanding of these models has been improving steadily. Of significant interest is the recent development of general theory and techniques for establishing population persistence \cite{hening2018,smith2011}. Even though the progress has been significant, there is currently a gap between the criteria that ensure persistence of species and the ones that ensure extinction in stochastic environments. In the deterministic setting this issue has been partly resolved using Morse decompositions of the extinction set \cite{hofbauer1989,schreiber2000}. An important open problem is whether one can do something similar in the stochastic setting. This would significantly increase the range of applications for the mathematical theory of stochastic persistence and extinction. 

One of the most important problems from ecology is to understand how species interactions and environmental fluctuations determine species distributions. There is strong evidence that spatial effects can change the way species interact as well as whether they survive or not. Sometimes the framework of reaction-diffusion equations is used to model spatial ecological processes \cite{cantrell2004}. The strength of these partial differential equation (PDE) models is that one can use information about how individuals move, die, interact or reproduce in order to gain knowledge regarding the coexistence of species. A serious limitation, however, has been the fact that most reaction-diffusion models studied in ecological contexts do not consider the effects of environmental fluctuations, although there is already a first influx of stochastic PDE theory into ecology \cite{kuehn2013}. There are many additional ways in which one can model fluctuations in this setting. Recently it has been shown that one can extend previous stochastic persistence results to a class of very general Markov processes \cite{benaim2018}. Nevertheless, these results do not work for many scenarios where one has continuous, spatial structure. 

There are many open problems which must be solved to improve our understanding of the mechanisms that make a complex ecosystem stable or unstable as well as the deterministic and stochastic fluctuations that can alter the stability of an ecosystem. Improving our understanding of how systems are stable/unstable as well as what causes systems to be stable/unstable will lead to improvements in forecasting when a system will change stability. In particular, improved understanding will likely lead to new and improved ``early warning signals'' of ecosystem collapse. 
    
These different lines of research have to a large extent been developed in a relatively independent manner with little interaction between ecologists, mathematicians, and physicists. From 08--12 August, 2022, 32 individuals participated in a workshop at the Lorentz Center, Leiden, The Netherlands in order to arrive at an improved understanding of stability, vulnerability, and other ecosystem properties. An interdisciplinary dialogue between these different groups provided a foundation of important problems to consider over the next 5-10 years. In particular, the workshop participants noted the importance of (1) improving our understanding of the effect of scale, both temporal and spatial, for both deterministic and stochastic problems, and including the role of species versus functional groups; (2) clarifying the different terminologies and definitions used in different scientific fields, noting that there are differences of thought with regards to “stability”, a main theme of the workshop; (3) developing a comprehensive set of data analysis techniques arising from different fields but which can be used together to improve our understanding of existing data sets; (4) having theoreticians/computational scientists collaborate closely with empirical ecologists to determine what new data should be collected, including data that will improve our understanding of the different types of noise in data; (5) improving our knowledge of how to protect and/or restore ecosystems by understanding when transitions occur and if they are good or bad, smoothing or extending transitions, and using feedback to control transitions; (6) incorporating socio-economic effects into models of ecosystems; (7) improving our understanding of the role of deterministic and stochastic fluctuations; (8) studying the current state of biodiversity at the functional level, taxa level and genome level and improving our knowledge of how biodiversity may change in the future and how it may be maintained. By focusing on the above areas of interest, researchers can build momentum to fundamentally improve our understanding of complex biological communities and how these communities will be affected by demographic, environmental, and seasonal change and disturbance.

\section{The Role of Ecological Modelling}
Before providing detail regarding the problems outlined above, it is important to note that modelling plays a critical role in each of the eight areas of interest. At a base level, one should consider the purpose of modelling. Should models be used as a mechanism for generating hypotheses; should they be used for making predictions; should models be guided by theory or by observational data. Moreover, it is worth considering if it is at all realistic to model an ecological system with many thousands of components. Unlike in physics-based models which have well-defined forces and scales over which different factors must be incorporated, it is difficult to know what should be included in ecology-based models. There is also a danger in modelling with too many parameters, thereby creating layers of complexity which may not be fully understood - sensitivity analysis may reveal which parts of a model are not worth making more complex. To ameliorate these issues, one can base models on ecologists’ foundational understanding of the system in question, formulate models for very specific questions that need answering, and incorporate model verification via laboratory or field experimentation. One might also consider using temporal and spatial data directly to answer a question rather than developing a mathematical model. Yet another possibility would be to use machine learning to derive or select a model based on data. Although a machine learning-based model could provide some predictive power, it would not provide much insight into the underlying ecological processes. It is well-worth remembering the value, and the possible pitfalls, of modelling complex ecosystems as we now consider the eight problem areas of interest in detail.
\clearpage
\section{Important Problems to Consider}\label{sec:problems}

\vspace{5mm}

\noindent {\bf (1) Improve our understanding of the effect of scale, both temporal and spatial, for both deterministic and stochastic problems, and including the role of species versus functional groups.}\\

Within ecology, there is a long history of theoretical and empirical studies to understand complex ecosystems. Much of this work is predicated on an assumption that the systems exist at an asymptotically stable equilibrium. Additionally, much of the work is based on considering interactions between individual species or, even more broadly, functional groups. However, recent work by Rooney, et al. \cite{rooney2008} has demonstrated the importance of understanding the effect of scale and the types of dynamics that are observed in real systems.

In particular, it seems to be important to relate individual level behavior to population level properties in order to understand how different types of individual behavior will affect the population dynamics and ecosystem functions. A major challenge involves the creation of simple models which incorporate the different scales, from individual to population level, and which incorporate fluctuations, deterministic and stochastic, at the different scales. It is worth noting that the level of the mean field is not the same for all complex models. For example, microbes are more different from each other than predators are from each other, and one should account for different interactions at different scales. At yet another level of scale, one can study how changes at the trait level affect dynamics and function at the population level. Arguably, trait diversity is the most important kind of diversity.

Scale is an important factor, along with the type of model, kind of data, and specific question, in determining if ecological models are good enough for making accurate predictions. Two examples of well-calibrated models that give decent estimates include models for CO$_2$ forecasting in soil systems, and models used by fisheries to predict fish populations. Other models can give general principles, but no accurate predictions. It largely depends on the granularity/scale of the models. 

When considering different scales, one can think about issues involving temporal scales, spatial scales, and the number of individuals in the system.  For temporal scales, if one is focusing on environmental drivers, it can be difficult to elicit information about the time scales of  important variation (e.g., daily and seasonal temperature changes, and longer term climate changes). Note that diurnal fluctuations are often ignored in models, and moreover, data is often unavailable or unclear for whether diurnal changes even matter.  Importantly, if there is a separation of time scales, then some scales of variation could be ignored, thus leading to a simpler model. For spatial scales, one could consider using metapopulation models. With regards to spatial patches, one should consider the scale of interest and whether the system is self-organized (spatial structure emerges from the interactions) or not (in which case external factors drive the patchiness, and fragmentation may be regarded as a stressor). For models, say on carbon fluxes, one needs spatial data, e.g., meta-ecosystems. Lots of data is available via remote sensing, but the data is not always at the appropriate resolution or scale.  And in considering the number of individuals, it is possible to couple a continuous ordinary differential equation (ODE) model for large populations to a discrete, stochastic model for small populations \cite{vaesen2019}, switching between the two depending on the population size.  Both demographic and environmental stochasticity are important in small populations, while environmental stochasticity is the main source of stochasticity for large populations.

With respect to temporal scale, note that long-term transients that are near an unstable steady state can appear to be stable on the observational time scale \cite{vanOpheusden2015}. Beyond this, one might consider whether it is even possible to distinguish systems with long-term transient states from systems in which one is driven from a state to another state by external forcing. Similarly, one could consider whether it is possible to distinguish oscillations from random fluctuations. Again, the challenge is to make predictions on the relevant scale. With regards to stochasticity, white noise is often applied as a small parameter in ecological models, but there is often large, non-white noise in real ecological systems. 

We should infer processes from patterns as well as the other way around. To move forward, it is important to obtain data on the processes (rather than patterns). To accomplish this, we should think about the types of data that must be collected to better understand general ecological principles. As one example for soil food webs, one might consider an ideal data set to be one for different soil types with measurements of the stocks and fluxes. One could use the data to determine which group of organisms is related to which type of processes. This would reduce the complexity in data and allow one to perform an improved analysis. Returning to the issue of scale, it might be difficult to identify these interactions at the species level, but it should be possible at higher organizational levels, such as functional groups.\\

\noindent {\bf (2) Clarify the different terminologies and definitions used in different scientific fields, noting that there are differences of thought with regards to ``stability''.}\\

Throughout the workshop, it was noted, in presentations and in discussion, that different scientific fields can use different terminologies for the same concept or different definitions for the same terminology. Additionally, it can happen that historical work is lost and rediscovered or rebranded. In particular, there seems to be a need to clearly define the “stability” of a system. One possibility is to consider the stability of coexistence as the ability for each species to invade the system, while another is to directly consider the stability of a coexistence state.  The former is based on local stability analysis of one or more extinct states, and the latter on local stability analysis of a coexistence state. More generally, one might consider different motivations for measuring stability including stabilizing productivity, species diversity, or a vulnerable species.

Furthermore, we may want to distinguish between approaches that reveal the deep structure of a network or system (e.g., positive and negative feedbacks, fluxes) and approaches that only remain on the surface of the network (e.g., local stability analysis). Hence, it may be important to not only look at whether a system is stable or unstable but also at how the energy is flowing through a network and to detect the forces (i.e., positive and negative feedbacks) that tend to keep the system together or lead to its degradation. Related to stability, we note that already there are multiple definitions of resilience in use today \cite{krakovska2021}. Because ecology is a complex and wide-spanning field, a unique definition for various concepts likely is not practical. However, it would be useful to develop a glossary of terms related to stability and resilience along with definitions and descriptions of the relations between each definition or metric.

Two approaches based on linearisation (i.e., local stability analysis) mentioned above involve invasibility (local linearisation around an extinction state) versus persistent stability (local linearisation around the coexistence state). The linearisation process is already a simplification of the complicated dynamics, and yet, there remain a number of important questions/problems to study with respect to these two approaches. To start, what are the relationships between the two approaches, and is it possible to find examples where the analyses propose different answers. And if so, what does it mean, especially since both approaches are based on the same theory. Is there a way to bridge the two perspectives, and is there a way to perform a form of global analysis. 

Other important questions to consider with regards to stability of complex ecosystems are related to the sizes of ecosystems. Often, one considers small modules (2-4 species) or very large/infinite numbers (using matrix theory). Real systems are probably somewhere in the middle, but we lack the technical tools to work on those intermediate-sized systems. Methods used to simplify these systems (e.g., random interactions, linearisation) are based on assumptions that don’t apply in real systems. For example, real systems have properties that are common and repeatable. We need to develop new methods that account for the properties (structure) of real systems. Related to the system size, to how many dimensions can we reduce complex ecological systems. And how appropriate is the pairwise approach for ecological networks. One bottleneck is that we have statistical issues of detecting in ecological data whether interactions are pairwise or higher-order. Tools are being developed in network science but the question seems unresolved as far as we know. 

More generally, as we gain a better understanding of the stability of complex networks, we will be able to assess instability in a wide range of natural systems from economics and mental health to climate change. In light of this, the focus should include understanding generic forces driving complex systems. Broadly, there is a need to transfer nonstationary dynamical systems theory from more theoretical mathematical settings to concrete applications for complex systems.\\

\noindent {\bf (3) Develop a comprehensive set of data analysis techniques arising from different fields but which can be used together to improve our understanding of existing data sets.}\\

There exist numerous ecological data sets which have been studied using specific statistical and data analysis techniques appropriate to a specific question. But it is likely that we can improve our understanding of many of these data sets if we revisit them and use a comprehensive set of data analysis methods arising from different fields. By developing a comprehensive set of techniques, including machine learning and topological data analysis, researchers will be better enabled to understand the collected data. 

Broadly, we would like to improve our use of existing data-driven theory to analyze and understand data. Given a variety of data, including time series data, temporal data, and spatial data, including snapshot data, what analytical tools must be developed or borrowed from other disciplines to elicit new knowledge and information. What are the goals for analyzing the data: forecasting, general statistical analysis, ruling “out” models, which is easier than ruling “in” models, etc. If one considers a global approach to dynamics, it is of interest to understand causation, and to recover interactions via joint changes in time, along with understanding global dynamics and estimating the dimension of the system. To develop a comprehensive set of tools, we must understand how the tools and approaches vary across disciplines. In addition, since we currently have a significant amount of data, machine learning tools could be used to tell us if there is anything to extract from the data. 

One general approach worth considering is nonparametric forecasting, in which one uses data without having to explicitly build a model.  The data tells one the dynamics of the system, and enables one to predict the next few values in time. One might consider the relevance of statistics for understanding causation of interactions, as well as the use of delay embedding. Note however, if the model is fundamentally discontinuous, then one cannot use a delay reconstruction approach. Moreover, delay reconstruction largely does not provide information about the mechanisms driving the system dynamics, although the Sugihara group is working on understanding causation using delay embedding \cite{ye2015}.

In developing a coherent set of methods, it is also important to think about the kind of data we should be gathering in the future. In particular, we should obtain data on the ecological processes rather than merely the ecological patterns. General ecological theory seems to be going out of fashion. We should now be obtaining and analyzing data to discern general principles.

There may be a role for machine learning/artificial intelligence in ecology. While machine learning won’t replace our understanding of ecological processes, it could be a good tool for hypothesis generation.  If machine learning can find information then at least you know that the information is in the data, and it provides a reason to explore mechanisms. It may also be useful for making predictions of how a system will evolve, albeit without providing information about why the system is evolving in a particular way. 

Machine learning can be used in different ways, supervised and unsupervised. Some approaches could be based on data alone, while other approaches could fuse data with models \cite{jiahao2021,qraitem2020}. The models could be well-accepted and the machine learning is providing good model parameterisation, or if a model is not known, machine learning can help to develop an appropriate form of the model. Also, machine learning could be used for more targeted questions such as parameter estimation from data, feature selection, and determination of the best summary statistics for fitting data to models.

Much work has been performed using network/graph theory and topological data analysis in other fields, and it may be useful to adapt these ideas to improve our understanding of ecological problems. While graph/network theory has been used a lot in ecology, often there are no fixed network structures for real ecological systems. Instead, one must consider adaptive networks. In either scenario, these networks generally have a very high dimension and it is not clear how one can reduce the dimension in order to perform mathematical analysis. The field of power grid analysis may shed light on how the vulnerability of networks depends on connectivity/nodes/edges, etc. A lot of theory has been developed for networks that are time-dependent and which incorporate rewiring; this theory may be useful in studying ecological systems.

To develop a comprehensive set of techniques that can be used to analyze data and address the gaps between models and data, one should understand how to bridge the data-model gap and should have some comprehensive examples of how data analysis tools have been used successfully and unsuccessfully. To bridge the data-model gap, one uses expert knowledge and data (real or simulated) to develop a model and answer questions. Data collection, data analysis, model building and developing new research questions are not hierarchical. Rather, these steps form a cycle, where each phase feeds into the next phase, and the results produce new questions.  We now understand some systems very well (soil carbon systems, pollinator systems) because of the successful implementation of the data-model cyclical process. One should also understand negative results which may arise from the modelling process. Also, the methods must be categorized as local versus global, parametric versus nonparametric, etc., with ``rules'' for when to use which tool and the pitfalls to avoid. \\

\noindent {\bf (4) Encourage theoreticians/computational scientists to collaborate closely with empirical ecologists to determine what new data should be collected, including data that will improve our understanding of the different types of noise in data.}\\

One aspect of data is associated with the noise within the data. The noise should not be discarded, and we should think about how to properly include noise into our models. We should think about how the amalgamation of data from different collections affects the noise profile of the entire data set. When considering noise, ideally one would build a model for the variation in the data, instead of averaging across it or even ignoring/excluding the noise. 

Because there are always more rare than abundant species, sampling is important and the sampling design will affect what is observed. The rare species, which may be immigrants that might or might not be able to invade the ecosystem, are actually quite critical for ecosystems (diversity, functional redundancy, insurance, etc.) so care must be taken to properly sample them. Overall, we felt that the emphasis does not always need to be on perfect data collection, which is impossible, but to instead model sources of variation and identify the largest sources of uncertainty.  Overall, incorporating structured error into models is promising and leads to the notion that one does not always have to collect better data.

In thinking about ecosystem/food web data, sometimes more detail is needed since it is often unclear which species actually interact since they may be spatially separated even if they co-occur at the resolution they were sampled.  In addition, it is often unknown which species are alive or dead, or active or dormant, in the existing data.  Furthermore, sometimes too much information may be summarized into one food web when in fact the interactions can change throughout a year.  Sampling at different times could be needed.  Theoretically, the idea of a blinking web of interactions, where in the course of time some species are not present because they are dormant, has not been studied much.    

Lastly, it would certainly be wonderful if we can develop methods that enable us to make a distinction between oscillations and random fluctuations. While some workshop participants were doubtful that it is at all possible, others pointed out that in biological control, success of intervention crucially depends on timing of intervention and that the oscillations there are quite well understood. \\

\noindent {\bf (5) Improve our knowledge of how to protect and/or restore ecosystems by understanding when transitions occur and if they are good or bad, smoothing or extending transitions, and using feedback to control transitions.}\\

In order to improve our capability of protecting and/or restoring ecosystems, we must improve our knowledge of transition or tipping points. Broadly, it is important to know when a system is approaching a tipping point since that knowledge may allow us to prevent the transition from happening in the first place. Even if we cannot prevent the transition (it may not be feasible to prevent it for a number of reasons, including political or socio-economic ones), we may still take appropriate action to minimize the damage if we know the transition is coming soon. Additionally, it may be possible to turn a catastrophic shift into a smoother transition which is easier to predict and to manage.

If we assume that tipping/transition points are worth investigating, we must determine 
which aspects of an ecosystem (e.g., biogeochemistry) are most relevant for predicting whether a system is in a “bad” state or heading in a “bad” direction. Given that there is much to investigate, and perhaps little time to do it in, we must also determine where to focus our efforts. Should we focus on theory development, developing tools for data analysis, or both of the above. If we think about a complex ecosystem like coral reefs, is it best to investigate the entire system as a whole, or focus on gaining a strong understanding of smaller parts of the system? Should we focus on preventing catastrophic shifts, or should we also start giving more attention to what we should do if we cannot prevent them? What can we do to restore systems that have already degenerated? We may want to focus on switching between two alternative stable states. Another interesting point might be to implement changes in the direction and amplitude of temperature into ecological models to study the impact of climate change on natural ecosystems, and how it affects tipping points.  To accomplish this, we should understand the effect of higher temperatures on different populations in the ecosystem. 

It is clear that we need to understand the ecological part first, but we also need to take human behavior into account. Even when we know the warning signs, and even if we know we are heading towards a catastrophic shift, it may still not be possible to prevent it because doing so generally will cost money. How can we include such factors in our models? Adaptation of species to the new conditions may be a factor in preventing ecosystem collapse (e.g., species that have lost their mutualists may be able to find new ones). This makes the eventual response of the ecosystem much more unpredictable, but can we use this information to improve the predictions of the model? The theory of eco-evolutionary dynamics has shown that evolution may strongly affect ecological dynamics. To what extent may rapid evolution also influence the expectation of critical transitions or tipping points in ecological systems under the current press perturbations that are associated with global change. Will they even give rise to novel early warning systems?

When thinking about restoring a system that has undergone a shift, what is needed to restore the system? Structurally, we must be able to predict the components of a food web which need rebuilding, or which need to be made more resilient to changing conditions (e.g., climate change). The intervention in complex systems is tricky because although we have available data, we are lacking analytical tools to make informed interventions.

In thinking about managing changing ecosystems, we have separated the topic into three parts where green color indicates methods currently available and red color indicates open areas of research (e.g. \cite{villa2015}).
\begin{itemize}
\item For a system undergoing a transition, how can one ensure a smooth transition and avoid hysteresis.
\begin{itemize}
\item \textcolor{ForestGreen}{Understand changes to the system via interactions and feedbacks.}
%\item \textcolor{ForestGreen}{Employ a local approach.}
\item \textcolor{BrickRed}{Develop new approaches which allow one to scale up to more species.}
\end{itemize}
\item How can one employ an intervention or a restoration of a system which has transitioned.
\begin{itemize}
\item \textcolor{ForestGreen}{Use of a pulse intervention, which is bounded in time so interactions/feedbacks do not change.}
%\item \textcolor{ForestGreen}{Employ a global approach}
\item \textcolor{BrickRed}{Perturb the system with ``optimal" spatial structures/direction in the network.}
\end{itemize}
\item How can one prevent collapse, increase resilience, and keep the system in a safe operating space. 
\begin{itemize}
\item \textcolor{ForestGreen}{Understand changes to the system (e.g., if one parameter is moving in a bad direction, can one compensate by changing a different parameter?).}
%\item \textcolor{ForestGreen}{Local approach}
\item \textcolor{BrickRed}{Develop new methods which allow one to scale up to more species and which enable one to find manageable/controllable parameters.}
\end{itemize}
\end{itemize}

When managing concrete ecosystems, one should first understand the system-specific dynamics, and then design system specific-methods. Lastly, given a list of available methods, one must determine the most appropriate method. In the above classification, we can think of complexity in terms of the number of variables (e.g., species) and the number of spatial dimensions.  Work is currently occurring on a few to many species, with major questions at the many-species level of how to apply control. For just a few species, there are questions about how to use spatial patterning in restoration.  Researchers mostly are not thinking about space in conjunction with a large number of species.

There is likely no universal solution to the problem. However, it is reasonable to develop a few general types of strategies for management that have a chance of actually working in real systems.  There are a few caveats.  The picture above is incomplete - one can have additional state variables emerge if new species are introduced, which could be bad (invasive species) or good (biological control).  Presumably all of the above approaches can include stochasticity, but little to no work has been performed with respect to managing noisy ecological systems.

It is important to note that a tipping point framework may not be useful for all of the very important questions ecologists might be looking at. The various definitions we use all have assumptions if discussed mathematically. While we keep the assumptions in mind as mathematicians, we do not always know if the assumptions are satisfied in the systems we study. Indeed, we don’t really need to expect all systems to be around an attractor and we often do not consider noise accurately (adding a Gaussian term) \cite{kuehn2022}. Thus, tipping points are not widely applicable for all ecological questions as they focus on the region of instability, while less sudden and drastic changes are often also relevant for ecosystems and their management. Lastly, we need to look carefully at the hard problems one is trying to solve in concert with the data that actually exists. A lack of data will severely hamper progress in this area. \\

\noindent {\bf (6) Incorporate socio-economic effects into models of ecosystems.}\\

Currently, ecosystem services define the processes and functions of ecosystems that are beneficial for humans. This human-centric perspective does not take into account the fact that humans have a responsibility to protect and promote a healthy state of ecosystems. In order to understand the specific tasks needed to sustain healthy ecosystems, we should reconsider the definition of the utility functions that are associated with ecosystem services. In particular, these utility functions may need to incorporate the human actions that need to be optimized to ensure a benefit for other species and ecosystems in general. Future ecosystem models should therefore incorporate socio-economic effects that enable one to explore how human beings can interact with ecosystems without harming them.

As mentioned in area (5) above, there are several types of questions one may be interested in when coupling ecological models with human behavior, including the prevention of an oncoming ecological catastrophe, adaptation of behavior after a catastrophe has occurred, or restoration of an ecosystem after catastrophe. Moreover, one might consider ways of adjusting the form of the catastrophe to lessen its impact. It is important to note that even if one knows a catastrophe is pending, it may not be feasible to prevent it because doing so could be very expensive financially. As we develop new models to incorporate socio-economic effects, we must learn the best ways to couple economic/social aspects to changing ecological systems. The models should include the societal/financial costs of different scenarios, including preventing catastrophe, recovering a system that is approaching catastrophe, adapting after a catastrophe, and adjusting the form of catastrophe. 

Although the incorporation of socio-economic aspects is clearly important and necessary to make meaningful adjustments in today’s changing world, the work is complicated by a myriad of factors. To start, we should consider smaller problems that relate to data and tools that are available. Citizen assemblies may serve as an interesting source of specific important problems as well as a data source. Citizen assemblies (small groups of about 100 individuals) presumably would also consist of individuals interested in working towards meaningful change. This approach often  happens in practice in the smallest community governing bodies, but the global ecological impact of local self-organization in communities is not well-studied. We could also consider specific systems, such as coral reefs, which have associated with them smaller networks of individuals interested in their survival (fisherman, scientists, and other stakeholders). These networks are self-selected and it should be easy to identify the most important individuals for implementing change. It is also critical that we listen to indigenous groups who have a keen understanding of their environment and surrounding ecosystems.

In considering how to couple socio-economic aspects to understanding and controlling ecological systems, one must consider how to incorporate the cost of ecosystem degradation in economic models to make change happen. One can provide compensation for destruction, but this is often not effective. Costs should be meaningful and should induce people to change their thinking about ecosystems. Time scales are also important. Dramatically changing a system will take a long time. Within the system as it currently exists, there should be ways to incorporate the way to attain a long time-scale change. As one example, economists suggest to tax companies that have large climate impact (i.e., the price of a plane ticket should incorporate CO$_2$ emission), so that the higher the cost, the lesser the use. However, many things must be taken into account including how specific policies may lead to discrimination against certain groups of individuals.

Instead of this top-down approach, one could also think about a bottom-up approach, beginning with the question of what drives individuals to do certain things. Individuals are certainly willing to make short-term changes that are not necessarily in their best interest, but it is not clear if large groups of individuals are willing to adopt these sorts of changes for long-term sustainability. In considering behavior, the actions that are being asked of individuals should be clear to individuals that the actions actually make a difference. If someone feels like there is nothing that can be achieved, it is easy to become demotivated. The issue of long time-scales mentioned above is difficult for humans to comprehend. It can even have counter-productive effects. For example, now that the harmful effects of flying are becoming more widely known, some people are choosing to fly even more as long as they still can. Lastly, governance scientists study power and discourse and have noted that one should be careful in changing individuals’ behavior due to unexpected follow-on effects. \\

\noindent {\bf (7) The role of deterministic and stochastic fluctuations.}\\

Deterministic fluctuations (e.g., seasonal forcing) and stochastic fluctuations (e.g., environmental or demographic noise) can cause intermittent oscillations and other fundamental changes to a system (e.g., switching between cyclic and stable dynamics, switching between two metastable states, or causing an extinction event from a metastable state). While work has been undertaken to understand all of these, success has been made in general for lower-dimensional systems. Another interesting problem which has only recently been studied is the role of eco-evolutionary conditions in affecting a system’s stability. 

A number of problems to study were suggested throughout the course of the workshop, and are listed below.

It is known that when a species goes extinct, there is a probabilistic likelihood associated with each path that a species can take towards extinction. The path with the highest likelihood is called the optimal extinction path. What can we learn from this concept with respect to species rescue and intervention strategies? 

Researchers often distinguish between demographic stochasticity and environmental stochasticity. Is it possible for noise to lie somewhere in between the two types of stochasticity? How does the noise scale with population size? And what is the role of temporal correlations in the noise? Moreover, how does one even measure the noise so that we can better implement stochastic effects in models?

In studying both deterministic/periodic and stochastic fluctuations, exactly what data needs to be collected from real systems, and how do fluctuations affect the data observed by ecologists?  It is important to note that everything is fluctuating in observations. We should focus on transients which depend on the noise (or even transients in the absence of noise) in addition to continued study of equilibria or limit cycles. Importantly, one should also consider if the system is always near an unstable equilibrium, or if there even is an equilibrium.

Additionally, there is a general interest in furthering our understanding of the role fluctuations play in questions of coexistence, stability, species dynamics, and persistence versus extinction. These notions are relevant to improving our understanding of biodiversity, discussed below. In particular, how do extinction and coexistence occur and how are they influenced by different parameter values. Although a complete classification of coexistence equilibria is an unsolved problem, it is known that one can have multiple coexistence states in a Lotka-Volterra system for a small number of species. \\

\noindent {\bf (8) Biodiversity.}\\

Over the years, a number of modelling approaches have been used to understand ecological systems and biodiversity. One popular approach involves random matrix models \cite{may2019}. However, these models are artificial and indicate the need to move on to more realistic structures.  Additionally, random matrix theory is difficult to test. Two better alternatives include modern coexistence theory, developed by Peter Chesson (see \cite{barabas2018} for a review), and Ellner et al.'s data-driven framework, which can assess underlying mechanisms for coexistence in a system \cite{ellner2019}. 

There are a number of subtleties associated with coexistence of a larger number of species, including storage effect, species-specific response to environment, strength of competition depending on the environment, buffering, and relative nonlinearity. In studying coexistence, since species interactions involve competing for resources, it is possible we should include resources in the model, noting that $N$ resources are needed if $N$ species are to coexist. With this in mind, it is worth pointing out that one can get $N$ species to coexist if there exist $N$ temporal niches, where each species is the best competitor during part of the year \cite{meyer2022}. In a spatially extended system, might the niches be spatiotemporal niches rather than spatial niches? Also note that random competition can lead to clustering of similar species that survive together \cite{scheffer2006}. It is also well-worth noting that most coexistence theory is for up to six species, and it is not clear how the theory can be scaled up to include more species.   

To improve understanding of biodiversity, we must bring data from the field closer to modelling and find ways for the two approaches to meet in a better way. The models should be used to guide the design of field studies. Statisticians should be involved in experimental design from the outset, and not after the data has been collected. Snapshot data is not the best data to understand dynamical systems. Sometimes spatial data contains temporal information, but one should be careful. Sometimes, a space-for-time substitution might work but we need additional information about how/when this works. 

There are many open questions related to biodiversity. Does field/experimental data inform us about real-world biodiversity and how to preserve it? How do we even show that biodiversity is good? In discussing the importance of biodiversity, it was pointed out that invasive species can preserve functions, but most ecologists do not like this as there seem to be only a few rational arguments for this. As one example, grass in intertidal areas not native to California replaced a native grass in San Francisco Bay. Now, an endangered bird likes the invasive grass. Does higher biodiversity lead to higher risk for humans (epidemiology)? As Vitousek et al. have pointed out, many of earth’s ecosystems are dominated by humans \cite{vitousek1997}.

Also, biodiversity is not in all cases an important measure to look at. In soil communities the species concept is quite unclear. Whenever we decide on taxonomic units (percent similarity in DNA) many arbitrary choices determine the number of species rather than the underlying biology. Interactions with bacteria lead to adaptive processes so immense that 97\% similarity might not make any sense. What is the assemblage of the community, and when are they active?

Biodiversity also occurs at different scales (individuals to ecosystems), with species somewhere in the middle. As scientists, we must choose the relevant scale, but most people still view biodiversity as species diversity. We have also moved to functional diversity, but most theoreticians are not considering updated models. Should functional groups be included in models? The fact that models are based on fixed nodes limits their applicability to the real world where everything is variable. To what extent can the idea be extrapolated to complex systems? Can we account for multi-stressors and things that are changing at the same time? Can we account for all complexity, and how robust are the measures?\\

\section{Final Thoughts}

In addition to the areas of interest outlined in Sec. \ref{sec:problems}, there were also brief discussions on two important issues. One of the discussions involved the importance of societal literacy. Environmental/societal issues require some level of awareness on those topics by the general public. Currently, the level of literacy is quite low. To achieve a higher level, education of youth at all school levels must be undertaken. We need to enable teachers to discuss the goals of modelling and the philosophical side of modelling. Qualitative aspects of modelling must be discussed. There is a great need, and also a great value, in this.

Another discussion involved cross-fertilisation, since this is required to solve large, complicated, important problems, such as those described throughout this paper. It is important to remember that when performing interdisciplinary work, the technical barriers are often smaller than the cultural barriers. It can take time before individuals are comfortable with each other's disciplines and the work can begin (there is a nice discussion of this aspect of interdisciplinary collaboration in the book Scale: The universal laws of life, growth, and death in organisms, cities, and companies, by Geoffrey West \cite{west2018}). 

This paper provides an overview of important problems in ecology as discussed during a recent workshop on Stability and Fluctuations in Complex Ecological Systems held at the Lorentz Center. In addition to providing a broad foundation, the paper provides a summary of possible directions to explore over the next 5-10 years. We hope that it is useful to the scientific community as we strive to improve our understanding of ecological systems.

%%%%%%%%%%%%%%%%%%%%%
% Acknowledgments
%%%%%%%%%%%%%%%%%%%%%
% You may wish to remove the Acknowledgments section while your paper 
% is under review (unless you wish to waive your anonymity under
% double-blind review) if the Acknowledgments reveal your identity.
% If you remove this section, you will need to add it back in to your
% final files after acceptance.

\section*{Acknowledgments}

Many thanks to the Lorentz Center (Leiden, The Netherlands) who hosted the workshop, Stability and Fluctuations in Complex Ecological Systems, from 08-12 August, 2022.

%\begin{appendices}

%\label{sec:resprate}

% Please reset counters for the appendix (thus normally figure A1, 
% figure A2, table A1, etc.).

% In certain cases, it may be appropriate to have a PRINT appendix in
% addition to (or instead of) an online appendix. In this case, please 
% name the print appendix Appendix A, and any subsequent appendixes (if 
% there are any) should be named Online Appendix B, Online Appendix C,
% etc.

% Counters for each appendix should match the letter of that appendix.
% For example, tables in Appendix C should be numbered table C1, table C2,
% etc. This applies to tables, equations, and figures.

% It's better not to use the \appendix command, because we have some
% formatting peculiarities that \appendix conflicts with.

%\renewcommand{\theequation}{A\arabic{equation}}
% redefine the command that creates the equation number.
%\renewcommand{\thetable}{A\arabic{table}}
%\setcounter{equation}{0}  % reset counter 
%\setcounter{figure}{0}
%\setcounter{table}{0}

%\end{appendices}

%appendix here

%%%%%%%%%%%%%%%%%%%%
% Bibliography
%%%%%%%%%%%%%%%%%%%%%

\renewcommand*{\bibfont}{\small}
\bibliographystyle{spbasic}	
\bibliography{Refs}

\begin{thebibliography}{35}
\providecommand{\natexlab}[1]{#1}
\providecommand{\url}[1]{{#1}}
\providecommand{\urlprefix}{URL }
\expandafter\ifx\csname urlstyle\endcsname\relax
  \providecommand{\doi}[1]{DOI~\discretionary{}{}{}#1}\else
  \providecommand{\doi}{DOI~\discretionary{}{}{}\begingroup
  \urlstyle{rm}\Url}\fi
\providecommand{\eprint}[2][]{\url{#2}}

\bibitem[{Barab{\'a}s et~al.(2018)Barab{\'a}s, D'Andrea, and
  Stump}]{barabas2018}
Barab{\'a}s G, D'Andrea R, Stump SM (2018) Chesson's coexistence theory.
  Ecological Monographs 88(3):277--303

\bibitem[{Benaim(2018)}]{benaim2018}
Benaim M (2018) Stochastic persistence. arXiv preprint arXiv:180608450

\bibitem[{Billings and Forgoston(2018)}]{billings2018}
Billings L, Forgoston E (2018) Seasonal forcing in stochastic epidemiology
  models. Ricerche di Matematica 67(1):27--47

\bibitem[{Cantrell and Cosner(2004)}]{cantrell2004}
Cantrell RS, Cosner C (2004) Spatial ecology via reaction-diffusion equations.
  John Wiley \& Sons

\bibitem[{De~Ruiter et~al.(1995)De~Ruiter, Neutel, and Moore}]{deRuiter1995}
De~Ruiter PC, Neutel AM, Moore JC (1995) Energetics, patterns of interaction
  strengths, and stability in real ecosystems. Science 269(5228):1257--1260

\bibitem[{Dougoud et~al.(2018)Dougoud, Vinckenbosch, Rohr, Bersier, and
  Mazza}]{dougoud2018}
Dougoud M, Vinckenbosch L, Rohr RP, Bersier LF, Mazza C (2018) The feasibility
  of equilibria in large ecosystems: A primary but neglected concept in the
  complexity-stability debate. PLoS Computational Biology 14(2):e1005988

\bibitem[{Ellner et~al.(2019)Ellner, Snyder, Adler, and Hooker}]{ellner2019}
Ellner SP, Snyder RE, Adler PB, Hooker G (2019) An expanded modern coexistence
  theory for empirical applications. Ecology Letters 22(1):3--18

\bibitem[{Forgoston and Moore(2018)}]{forgoston2018}
Forgoston E, Moore RO (2018) A primer on noise-induced transitions in applied
  dynamical systems. SIAM Review 60(4):969--1009

\bibitem[{Gardner and Ashby(1970)}]{gardner1970}
Gardner MR, Ashby WR (1970) Connectance of large dynamic (cybernetic) systems:
  critical values for stability. Nature 228(5273):784--784

\bibitem[{Hening and Nguyen(2018)}]{hening2018}
Hening A, Nguyen DH (2018) Coexistence and extinction for stochastic
  {K}olmogorov systems. Annals of Applied Probability 28(3):1893--1942

\bibitem[{Hofbauer and So(1989)}]{hofbauer1989}
Hofbauer J, So JWH (1989) Uniform persistence and repellors for maps.
  Proceedings of the American Mathematical Society 107(4):1137--1142

\bibitem[{Jacquet et~al.(2016)Jacquet, Moritz, Morissette, Legagneux, Massol,
  Archambault, and Gravel}]{jacquet2016}
Jacquet C, Moritz C, Morissette L, Legagneux P, Massol F, Archambault P, Gravel
  D (2016) No complexity--stability relationship in empirical ecosystems.
  Nature Communications 7(1):12573

\bibitem[{Jiahao et~al.(2021)Jiahao, Hsieh, and Forgoston}]{jiahao2021}
Jiahao TZ, Hsieh MA, Forgoston E (2021) Knowledge-based learning of nonlinear
  dynamics and chaos. Chaos: An Interdisciplinary Journal of Nonlinear Science
  31(11):111101

\bibitem[{Krakovsk{\'a} et~al.(2021)Krakovsk{\'a}, Kuehn, and
  Longo}]{krakovska2021}
Krakovsk{\'a} H, Kuehn C, Longo IP (2021) Resilience of dynamical systems.
  arXiv preprint arXiv:210510592

\bibitem[{Kuehn(2013)}]{kuehn2013}
Kuehn C (2013) Warning signs for wave speed transitions of noisy
  {F}isher--{KPP} invasion fronts. Theoretical Ecology 6:295--308

\bibitem[{Kuehn et~al.(2022)Kuehn, Lux, and Neam{\c{t}}u}]{kuehn2022}
Kuehn C, Lux K, Neam{\c{t}}u A (2022) Warning signs for non-{M}arkovian
  bifurcations: colour blindness and scaling laws. Proceedings of the Royal
  Society A 478(2259):20210740

\bibitem[{MacArthur(1955)}]{macarthur1955}
MacArthur R (1955) Fluctuations of animal populations and a measure of
  community stability. Ecology 36(3):533--536

\bibitem[{May(1972)}]{may1972}
May RM (1972) Will a large complex system be stable? Nature 238:413--414

\bibitem[{May(2019)}]{may2019}
May RM (2019) Stability and complexity in model ecosystems. Princeton
  University Press

\bibitem[{Meyer et~al.(2022)Meyer, Steinmetz, and Shnerb}]{meyer2022}
Meyer I, Steinmetz B, Shnerb NM (2022) How the storage effect and the number of
  temporal niches affect biodiversity in stochastic and seasonal environments.
  PLOS Computational Biology 18(3):e1009971

\bibitem[{Neutel and Thorne(2014)}]{neutel2014}
Neutel AM, Thorne MA (2014) Interaction strengths in balanced carbon cycles and
  the absence of a relation between ecosystem complexity and stability. Ecology
  Letters 17(6):651--661

\bibitem[{van Opheusden et~al.(2015)van Opheusden, Hemerik, van Opheusden, and
  van~der Werf}]{vanOpheusden2015}
van Opheusden JHJ, Hemerik L, van Opheusden M, van~der Werf W (2015)
  Competition for resources: complicated dynamics in the simple {T}ilman model.
  SpringerPlus 4:1--31

\bibitem[{Picoche and Barraquand(2019)}]{picoche2019}
Picoche C, Barraquand F (2019) How self-regulation, the storage effect, and
  their interaction contribute to coexistence in stochastic and seasonal
  environments. Theoretical Ecology 12:489--500

\bibitem[{Qraitem et~al.(2020)Qraitem, Kularatne, Forgoston, and
  Hsieh}]{qraitem2020}
Qraitem M, Kularatne D, Forgoston E, Hsieh MA (2020) Bridging the gap:
  {M}achine learning to resolve improperly modeled dynamics. Physica D:
  Nonlinear Phenomena 414:132736

\bibitem[{Rooney et~al.(2008)Rooney, McCann, and Moore}]{rooney2008}
Rooney N, McCann KS, Moore JC (2008) A landscape theory for food web
  architecture. Ecology Letters 11(8):867--881

\bibitem[{Scheffer and Van~Nes(2006)}]{scheffer2006}
Scheffer M, Van~Nes EH (2006) Self-organized similarity, the evolutionary
  emergence of groups of similar species. Proceedings of the National Academy
  of Sciences 103(16):6230--6235

\bibitem[{Schreiber(2000)}]{schreiber2000}
Schreiber SJ (2000) Criteria for $c^r$ robust permanence. Journal of
  Differential Equations 162(2):400--426

\bibitem[{Smith and Thieme(2011)}]{smith2011}
Smith HL, Thieme HR (2011) Dynamical systems and population persistence, vol
  118. American Mathematical Society

\bibitem[{Vaesen et~al.(2019)Vaesen, Scherjon, Hemerik, and
  Verpoorte}]{vaesen2019}
Vaesen K, Scherjon F, Hemerik L, Verpoorte A (2019) Inbreeding, {A}llee effects
  and stochasticity might be sufficient to account for {N}eanderthal
  extinction. PLoS One 14(11):e0225117

\bibitem[{van Velzen et~al.(2022)van Velzen, Gaedke, and
  Klauschies}]{vanVelzen2022}
van Velzen E, Gaedke U, Klauschies T (2022) Quantifying the capacity for
  contemporary trait changes to drive intermittent predator--prey cycles.
  Ecological Monographs 92(2):e1505

\bibitem[{Villa~Mart{\'\i}n et~al.(2015)Villa~Mart{\'\i}n, Bonachela, Levin,
  and Mu{\~n}oz}]{villa2015}
Villa~Mart{\'\i}n P, Bonachela JA, Levin SA, Mu{\~n}oz MA (2015) Eluding
  catastrophic shifts. Proceedings of the National Academy of Sciences
  112(15):E1828--E1836

\bibitem[{Vitousek et~al.(1997)Vitousek, Mooney, Lubchenco, and
  Melillo}]{vitousek1997}
Vitousek PM, Mooney HA, Lubchenco J, Melillo JM (1997) Human domination of
  {E}arth's ecosystems. Science 277(5325):494--499

\bibitem[{Wang et~al.(2019)Wang, Cadotte, Chen, Fraser, Zhang, Huang, Luo, Shi,
  and Loreau}]{wang2019}
Wang Y, Cadotte MW, Chen Y, Fraser LH, Zhang Y, Huang F, Luo S, Shi N, Loreau M
  (2019) Global evidence of positive biodiversity effects on spatial ecosystem
  stability in natural grasslands. Nature Communications 10(1):3207

\bibitem[{West(2018)}]{west2018}
West G (2018) Scale: The universal laws of life, growth, and death in
  organisms, cities, and companies. Penguin

\bibitem[{Ye et~al.(2015)Ye, Deyle, Gilarranz, and Sugihara}]{ye2015}
Ye H, Deyle ER, Gilarranz LJ, Sugihara G (2015) Distinguishing time-delayed
  causal interactions using convergent cross mapping. Scientific Reports
  5(1):1--9

\end{thebibliography}

%\newpage{}

%\section*{Tables}
%\renewcommand{\thetable}{\arabic{table}}
%\setcounter{table}{0}

\end{document}